\documentclass[aps,pre,reprint,10pt,twocolumn,nofootinbib,showpacs]{revtex4-1}
\usepackage{graphicx}
\usepackage{amssymb}
\usepackage{amsmath}
\usepackage{amsfonts}
\usepackage{color}
\usepackage[latin1]{inputenc}
\usepackage{float}

\begin{document}

\title{Time increasing rates of infiltration and reaction in porous media at the
percolation thresholds}

\author{Ismael S. S. Carrasco}
\email{theismiu@gmail.com}
\affiliation{Instituto de F\'\i sica, Universidade Federal Fluminense,\\
Avenida Litor\^anea s/n, 24210-340 Niter\'oi RJ, Brazil}
\author{F\'abio D. A. Aar\~ao Reis}
\email{fdaar@protonmail.com}
\affiliation{Instituto de F\'\i sica, Universidade Federal Fluminense,\\
Avenida Litor\^anea s/n, 24210-340 Niter\'oi RJ, Brazil}

\begin{abstract}

The infiltration of a solute in a fractal porous medium is usually anomalous,
but chemical reactions of the solute and that material may increase the porosity
and affect the evolution of the infiltration.
We study this problem in two- and three-dimensional lattices with randomly distributed
porous sites at the critical percolation thresholds and with a border in contact
with a reservoir of an aggressive solute.
The solute infiltrates that medium by diffusion and the reactions with the impermeable sites
produce new porous sites with a probability $r$,
which is proportional to the ratio of reaction and diffusion rates
at the scale of a lattice site.
Numerical simulations for $r\ll1$ show initial subdiffusive scaling and long time
Fickean scaling of the infiltrated volumes or areas, but with an intermediate regime
with time increasing rates of infiltration and reaction.
The anomalous exponent of the initial regime agrees with a relation previously applied to
infinitely ramified fractals.
We develop a scaling approach that explains the subsequent time increase of the infiltration
rate, the dependence of this rate on $r$, and the crossover to the Fickean regime.
The exponents of the scaling relations depend on the fractal dimensions of the
critical percolation clusters and on the dimensions of random walks in those clusters.
The time increase of the reaction rate is also justified by that reasoning.
As $r$ decreases, there is an increase in the number of time decades of the intermediate regime,
which suggests that the time increasing rates are more likely to be observed is slowly reacting systems.

\end{abstract}

\maketitle

\section{Introduction}
\label{intro}

When a porous medium is filled with a static fluid and an external surface is put in contact with
a reservoir of a mobile species with a different concentration, that species may infiltrate into
or out of that medium by diffusion.
This occurs, for instance, in the weathering of
geological materials \citep{navarre2009}, in the dispersion of contaminants
in soils \citep{youStoten2020}, and in the transport of radionuclides in nuclear waste containers
\citep{medved2019}.
A simple model for the diffusive infiltration without reactions considers lattice random walks of
(excluded volume) particles that start from a source at one border, as originally
proposed by Sapoval et al \citep{sapovalJPL1985} and
recently extended to fractal geometries \citep{reis2016}.
The infiltration length $I$ is defined as the
infiltrated volume per unit area of the exposed surface and is shown to scale as
\begin{equation}
I\sim t^n ,
\label{defngeral}
\end{equation}
where $n$ is termed infiltration exponent.
If the medium in $d$ dimensions is homogeneous and the source is a border of dimension $d-1$,
the infiltration is Fickean, with $n=1/2$,
as shown in models and experiments \citep{Voller_wrr,filipovitch,reisvoller2019}.
However, anomalous exponents ($n\neq1/2$) were already obtained in studies of moisture
infiltration in construction materials \citep{kuntz,lockington,elabd2015,delgado2016,elabd2020},
in several models of infiltration in regular fractals
\citep{Voller_wrr,reis2016,reisbolstervoller}, and
in the infiltration of glycerin in Hele-Shaw cells with fractal pore geometry
(a problem that is equivalent to diffusive infiltration) \citep{filipovitch}.

These processes are related to the molecular diffusion in the medium.
When particles randomly move in a homogeneous medium, their
root-mean-square displacement ${\cal R}$ increases in time as $t^{1/2}$
(i.e. Fickean diffusion).
In highly disordered media, anomalous diffusion is observed, in which
\begin{equation}
{\cal R} \sim t^{1/d_W} ,
\label{defdW}
\end{equation}
with the random walk dimension $d_W\neq 2$ \citep{bouchaud,Havlin,metzler2014,fao2019}.
In fractal porous media, the self-similar distributions of irregularities, such as
impenetrable barriers or dead ends, usually lead to subdiffusion, in which $d_W>2$.
However, note that the infiltration exponent $n$ and the random walk exponent $1/d_W$ may be different.
For instance, in infinitely ramified fractals such as Sierpinski carpets and Menger sponges,
the connection of diffusion driven infiltration and the diffusion anomalies
led to a relation between $n$, $d_W$, the fractal dimension $d_F$ of the medium,
and the dimension $d_B$ of the infiltration border \citep{reis2016,reisvoller2019}.
A possible consequence is that superdiffusive infiltration ($1/2<n\leq1$) occurs in
a medium where random walks are subdiffusive \citep{reisbolstervoller}.

The interplay of infiltration in disordered media and chemical reactions
may lead to changes in the structure of those media, which is
of particular importance in the evolution of geological materials
\citep{noiriel2009,steefelRevMinGeo2015,noirielRevMinGeo2015,seigneurRevMinGeo2019}.
Diffusion is expected to be the dominant transport mechanism in several cases,
particularly in low porosity media \citep{sak2004,navarre2009,chagneau2015,guSci2020}.
For instance, in olivine-rich rocks whose porosity is near
a few percent, serpentinization creates small fractures that serve as pathways for fluid infiltration
\citep{jamtveit2008,tutolo2016,schwarzenbach2016}.
Moreover, water infiltration and loss of reaction products in basalt clasts lead to the formation of
porous weathered rinds around (more compact) unaltered cores
\citep{navarre2009,navarre2013}.
Fractal pore geometry was confirmed in both altered and unaltered domains of those clasts
\citep{navarre2013} and was also observed in many other geological materials
\citep{farin,sahimi1993}.
Infiltration anomalies are expected due to the fractality,
but they may change with the progress of the reactions.
Such changes were recently shown in a model of infiltration and reaction in infinitely ramified
fractals, in which the initial subdiffusive behavior crossed over to a long time
Fickan infiltration \citep{reisAWR2019}.

This scenario motivates the present investigation of the coupling of diffusive infiltration and
dissolution reactions (which create new porosity) in critical percolation clusters 
\citep{stauffer,sahimi1993}, which are the most prominent stochastic fractal media exhibiting
anomalous diffusion.
Our first step is to study the non-reactive infiltration of a solute
in hypercubic lattices of dimensions $d=2$ and $3$ where the porous sites are randomly distributed
with the critical percolation probabilities and the remaining sites are impermeable.
The infiltrated area ($d=2$) and volume ($d=3$) are shown to increase with exponents
$n<1/2$ consistent with the same scaling relation obtained in infinitely ramified fractals
\citep{reis2016}.
The second part of this work considers that reactions between
the infiltrating solute and the impermeable solid form new porous sites in which solute transport 
is possible.
The initial subdiffusive infiltration and the long time Fickean behavior are observed,
but they are separated by a regime in which the infiltrated and the dissolved volumes (or areas)
increase faster than linearly in time, i.e. infiltration and dissolution have time increasing rates.
This phenomenon results from the combination of a slow penetration of the solute in a
preferential direction of a porous system of vanishing density and the linear
advance in all directions of the dissolution of the solid walls surrounding the pores.
The time range of this regime is shown to increase as the reaction
rate decreases (relatively to the diffusion rate), so the nontrivial anomaly may last for long times
in slowly reacting materials with fractal pore systems.
These results are obtained with numerical simulations and explained by a scaling approach.

This paper is organized as follows.
Sec. \ref{model} presents the model of infiltration and reaction, information on the methods
of solution, and the main quantities to be measured.
Sec. \ref{review} reviews the results for the same model with an initially compact (not porous) medium,
which helps the interpretation of some results in the porous media.
Sec. \ref{infiltrationonly} presents results for infiltration without reactions in $d=2$ and $d=3$.
Sec. \ref{simulations} presents simulation results for the model with reactions.
Sec. \ref{scaling} presents a scaling approach that explains the observed evolution of
infiltration and reaction lengths, with particular emphasis on the regime with time increasing rates.
Sec. \ref{conclusion} summarizes our results and presents our conclusions.

\section{Models and Methods}
\label{model}

\subsection{Model Definition}
\label{modeldissolution}

The porous media are built in the region $z>0$ of hypercubic lattices of dimension $d$.
Each site is expected to represent a homogeneous mesoscopic region of a porous material.
The permeable sites, which are termed P sites, are inert.
Their initial fraction is equal to the percolation
threshold $p_c$ with nearest neighbor (NN) connectivity.
The impermeable sites contain a reactive material and are termed M sites;
their initial fraction is $1-p_c$.
With these definitions, solute molecules can be transported only through P sites that are NN.

\begin{figure*}[!ht]
\center
\includegraphics[clip,width=0.75\textwidth]{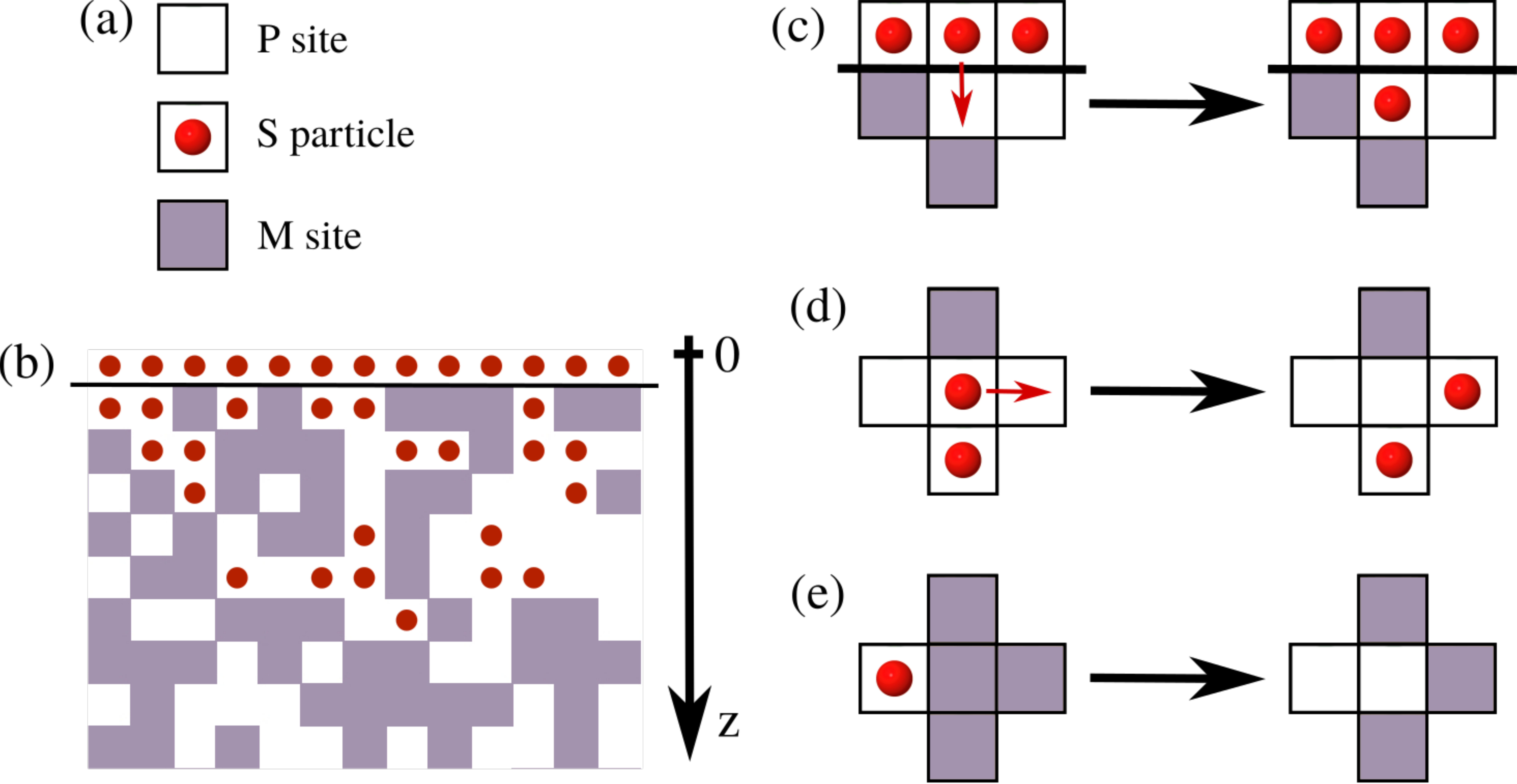}
\caption{(Color online) 
Illustration of the model in $d=2$.  (a) Types of sites and particles. 
(b) The porous medium with some S particles and with a border in contact with their source.
(c) Infiltration of an S particle from the border .
(d) Hop of an S particle to an empty P site at the right. 
(e) Reaction of an S particle with an M site.
}
\label{medium}
\end{figure*}

The pore solution and the material M are initially in chemical equilibrium.
A large solution at $z\leq0$ with a constant concentration of an aggressive solute is 
put in contact with the medium at time $t=0$.
Thus, $z=0$ is the infiltration border, whose dimension is $d_B=d-1$.
The solute is represented by S particles that permanently fill all the sites with $z=0$
and that may also occupy P sites of the medium.
Excluded volume conditions are considered for the S particles, so that a P site or a border site
may have zero or one S particle.
The possible configurations of the lattice sites are shown in Fig. \ref{medium}(a).
Figure \ref{medium}(b) shows a configuration of a lattice in $d=2$ after infiltration of some
S particles and the filled border at $z=0$.

In a time interval $\tau$, all S particles attempt to hop to a randomly chosen NN site.
If an S particle attempts to hop from the border $z=0$ to an empty P site, the hop is executed
and another S particle is immediately inserted at its initial position, so that
the border remains with the same solute
concentration; this is illustrated in Fig. \ref{medium}(c).
However, if the S particle attempts to hop to an M site or to a site with another S particle,
the attempt is rejected and that particle remains in the same position.
Fig. \ref{medium}(d) illustrates the hop of an S particle at a P site, which was executed
because the target site was an empty P site; however, if that particle attempted to hop
to the site below it or to the site above it, the attempt would be rejected.

In the same time interval $\tau$, an S particle in contact with a NN M site can react with
probability $r$.
The reaction leads to the transformation of the M site into an empty P site and to the
annihilation of the S particle, as illustrated in Fig. \ref{medium}(e).
This rule of the model is a simplified description of a series of physical and chemical processes:
(i) the reaction between S and M forms soluble and non-soluble products;
(ii) the non-soluble products form a porous precipitate with a
physical structure similar to those of the initial P sites;
(iii) chemical equilibrium is restored in the pore solution where the reaction occurs
(this is represented by the annihilation of the S particle).

\subsection{Interpretation of the model parameters}
\label{modelparameters}

Here we partly follow the reasoning of Ref. \protect\cite{reisAWR2019} to relate the model parameters
in $d=3$ to measurable quantities.

Letting $a$ be the lattice constant and $D$ be the diffusion coefficient of the S particles in
the porous medium formed only by P sites, we have
\begin{equation}
D = \frac{a^2}{2d\tau} .
\label{defD}
\end{equation}
This coefficient is expected to be smaller than that in free solution.

When an M site reacts, the change in the number of moles of the reacting material is
${\Delta n}_M=a^3/v$, where $v$ is the molar volume of that material;
in an application, $v$ depends on the properties of the reacting material and its volume
fraction in the solid represented by the M site.
The area of the M site in contact with a NN P site is $a^2$, but
the P site represents a mesoscopic porous region, so only a fraction of that area is in contact
with pore solution.
We assume that this fraction is equal to the effective porosity of a P site, which is denoted
as $\phi_P$; see e.g. Ref. \protect\cite{reisbrantley2019}.
Thus, an area $A_{\text M}=\phi_Pa^2$ of the M site is in contact with the solution in the NN P site.
Since the ${\Delta n}_M$ moles of M react with a probability $r$ in a time interval $\tau$,
the reaction rate $k$, in mol/(m${}^2$s), is given by
$k=r{\Delta n}_M/\left(A_M \tau\right) = 6rD/\left(av\phi_P\right)$;
here Eq. (\ref{defD}) was used with $d=3$.
This gives
\begin{equation}
r = \frac{\phi_P v}{6} \frac{ak}{D} .
\label{pkD}
\end{equation}

Thus, $r$ is a ratio between rates of reaction and diffusion in the volume of a lattice site,
which may be interpreted as a second Damkohler number of the model \citep{fogler}.
The condition of slow reaction compared to diffusion implies $r\ll 1$, which 
is considered throughout this work.
Since our main results are obtained in terms of the parameter $r$,
the application of the model to a real system is possible if
the physical and chemical parameters in Eq. (\ref{pkD}) are estimated.

\subsection{Methods of Solution}
\label{simulation}

Scaling relations are well known for the structural properties
of percolation clusters and the diffusion of tracers in those media
\citep{havlin1983b,stauffer,Havlin}.
Relations between infiltration and diffusion exponents in regular fractals are also known
\citep{reis2016,reisvoller2019} and can be tested in the infiltration model without
reactions at $p=p_c$.
Moreover, scaling approaches for the interplay of diffusion and
alteration reactions (interfacial dissolution reprecipitation mechanism \citep{hellmann2012})
were developed in Ref. \protect\cite{alteredlayer} for non-porous media;
an analogous approach was used in the study of the growth of passive layers on metallic surfaces \citep{passive2}.
A combination of these methods is used here to explain the scaling regimes of infiltration and reaction.

We also use kinetic Monte Carlo simulations to support the predictions of this scaling approach.
In $d=2$, we consider lattices with infiltration border ($z=0$) of size $L=1024a$, with very
large length in the $z$ direction, and with periodic boundary conditions in the other direction.
Some simulations are also performed in $L=2048a$ to discard the possibility of finite-size effects.
The initial fraction of pore sites is $p_c=0.59274605$ \citep{feng2008},
the values of $r$ range from ${10}^{-2}$ to ${10}^{-6}$, and
$100$ realizations are used to calculate average quantities, up to
the maximal simulation time ${10}^{7}\tau$.
In $d=3$, we consider lattices with infiltration border of size $L=256a$, with very
large length in the $z$ direction, and with periodic boundary conditions in the other directions.
Some simulations are also performed in $L=512a$ to discard the possibility of finite-size effects.
The initial fraction of pore sites is $p_c=0.3116060$ \citep{koza2016},
the values of $r$ range from ${10}^{-2}$ to ${10}^{-5}$, and $50{\text{--}}100$
realizations are used to calculate average quantities.
The maximal simulation time is ${10}^{5}\tau$ for most samples, but
${10}^{6}\tau$ for the smallest $r$.

A medium at $p_c$ has a percolating cluster of P sites
with fractal dimension $d_F=91/48$ in $d=2$ \citep{dennijs1979,nienhuis1982}
and $d_F=2.5230\pm0.0001$ in $d=3$ \citep{ballesteros1999}.
The random walk dimension in critical percolation clusters in $d=2$ obtained from
simulations is $d_W=2.8784\pm0.0008$ \citep{grassberger1999}.
In $d=3$, the scaling of the conductivity in percolation clusters
gives $d_W=3.806\pm0.001$ \citep{ballesteros1999,normand1995,kozlov2010}.
These values are used to test the scaling relations determined here.

\subsection{Basic Quantities}
\label{quantities}

The infiltration time is denoted as $t$ and a dimensionless infiltration time is defined as
\begin{equation}
T=\frac{t}{\tau} .
\label{deft}
\end{equation}

The number of infiltrated sites, i.e. P sites with a particle S, is denoted as $N_I$.
For a unified description in all spatial dimensions, we define a dimensionless infiltration
length $I$ as the ratio between the number of infiltrated sites, $N_I$, and the number of sites of
the infiltration border, ${\left(L/a\right)}^d$:
\begin{equation}
I=\frac{N_I}{{\left(L/a\right)}^d} .
\label{defI}
\end{equation}
From this definition, $Ia$ can be interpreted as the average length in the $z$ direction occupied
by the (non-contiguous) infiltrated P sites.
The part of the medium with distance $\leq Ia$ from the infiltration border
is termed the infiltrated region; it comprises most sites with S particles.

The number of M sites transformed into P sites (i.e. M sites that react) is denoted as $N_R$.
The dimensionless reaction length $R$ is defined as the ratio between this number
and the number of sites of the infiltration border, ${\left(L/a\right)}^d$:
\begin{equation}
R=\frac{N_R}{{\left(L/a\right)}^d} .
\label{defR}
\end{equation}
The length $Ra$ measures the extent of penetration of the reaction in the $z$ direction.

\section{Review of the Model with Compact Reactive Media}
\label{review}

\begin{figure*}[!ht]
	\center
	\includegraphics[clip,width=0.9\textwidth]{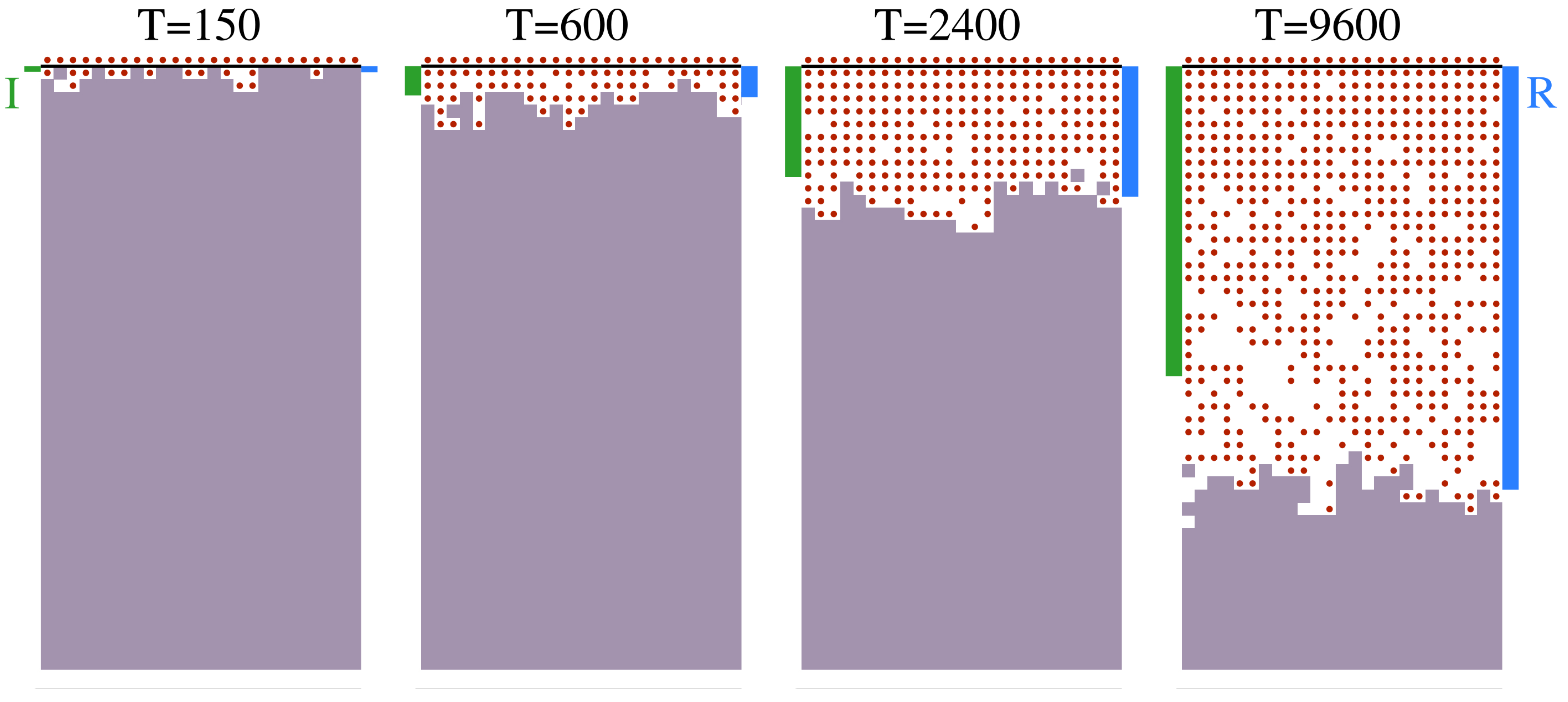}
	\caption{(Color online) Infiltration in a two dimensional compact medium with $r=10^{-2}$.
	The left (green) bar and the right (blue) bar represent the infiltration and reaction lengths,
	respectively.
	}
	\label{compact_perfil}
\end{figure*}

Here we consider the case in which the medium interacting with the solution is compact, i.e.
initially all lattice sites with $z>0$ are M sites.
Figure \ref{compact_perfil} shows snapshots of the attacked solid and the infiltrated region at
several times for $r={10}^{-2}$.

\begin{figure}[!ht]
\center
\includegraphics[clip,width=0.35\textwidth]{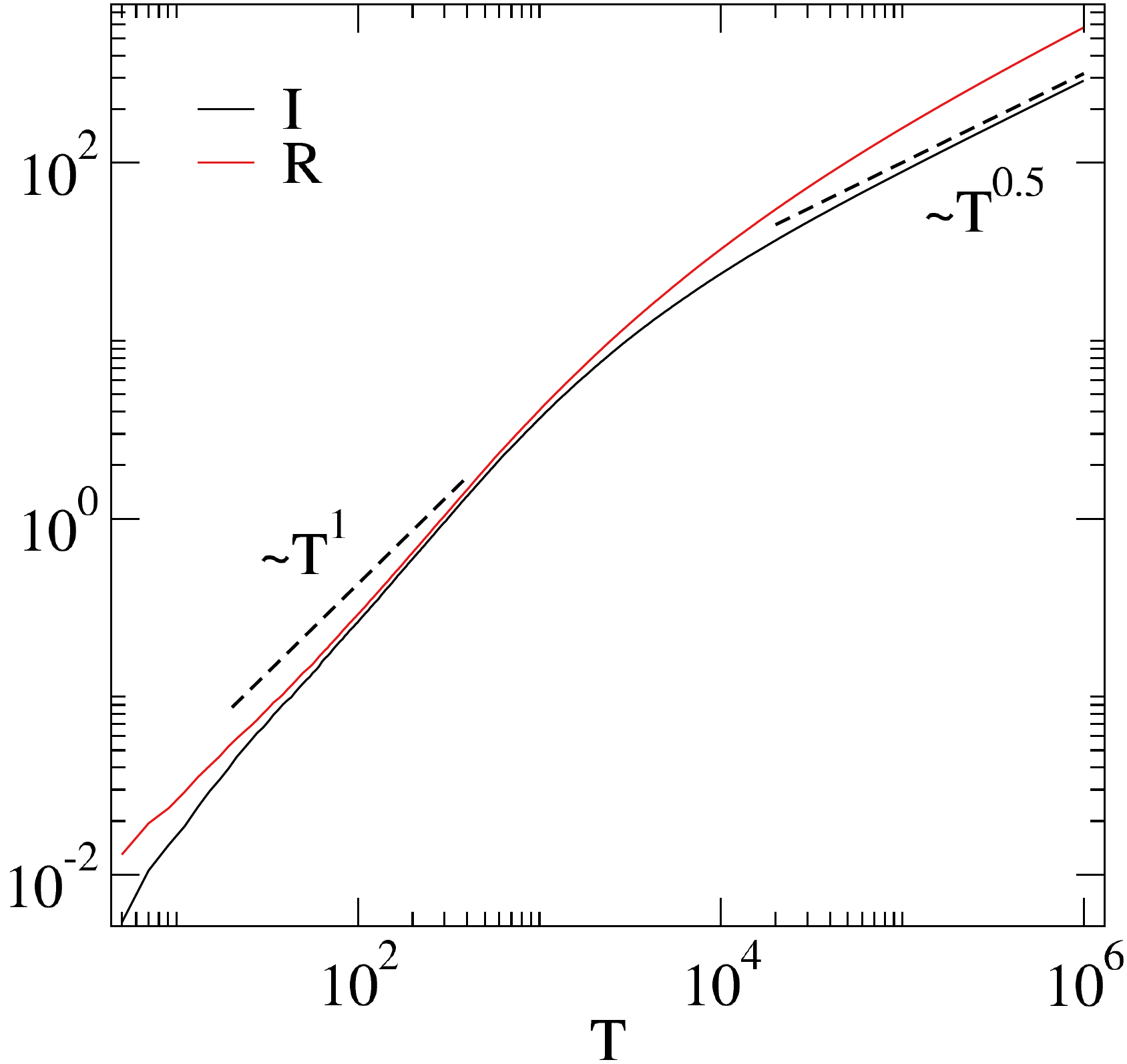}
\caption{(Color online)
Evolution of the infiltration and reaction lengths with an initially compact medium in $d=2$ and
$r=10^{-2}$.
}
\label{compact}
\end{figure}

In the shortest times, we observe a large density of S particles near the M surface because
they can rapidly
move in the porous medium near the source, whereas their reactions with M sites are relatively slow.
Thus, the M sites at the surface are almost all the time in contact with an S particle
(e.g. in the first three panels of Fig. \ref{compact_perfil}).
This is a regime controlled by the reaction, in which $I\sim rT$,
as shown in Fig. \ref{compact}.
However, when the average distance between the source and the M surface becomes large,
the time for a new particle to leave the source and to reach that surface increases.
This time will eventually be larger than the average time ($r^{-1}$) for a reaction with an M site,
which leads to a depletion in the density of S near the M surface.
In this condition, the infiltration is controlled by the diffusion of the S particles across the length $I$,
so $I\sim T^{1/2}$ ($R$ exceeds $I$ because the S particles occupy only a fraction of the sites that
reacted).
This is the long time Fickean regime, as shown in Fig. \ref{compact}.

The crossover time between the reactive and the Fickean regime is obtained by matching the expressions
for the infiltration length in those regimes:
\begin{equation}
T_c \sim r^{-2} .
\label{Tc}
\end{equation}
The right panel of Fig. \ref{compact_perfil}, which is at $T\approx T_c\sim{10}^4$, actually shows that the
density of S particles near the M surface is much smaller than the density at the shortest times.
Fig. \ref{compact} confirms that the infiltration is entering the Fickean regime at that time.

For further comparison, note that the crossover shown in Fig. \ref{compact} occurs with
a continuous decrease of the slope of the $\log{I}\times\log{T}$ and $\log{R}\times\log{T}$ plots.

\section{Infiltration without Reactions}
\label{infiltrationonly}

\begin{figure*}[!ht]
\center
\includegraphics[clip,width=0.9\textwidth]{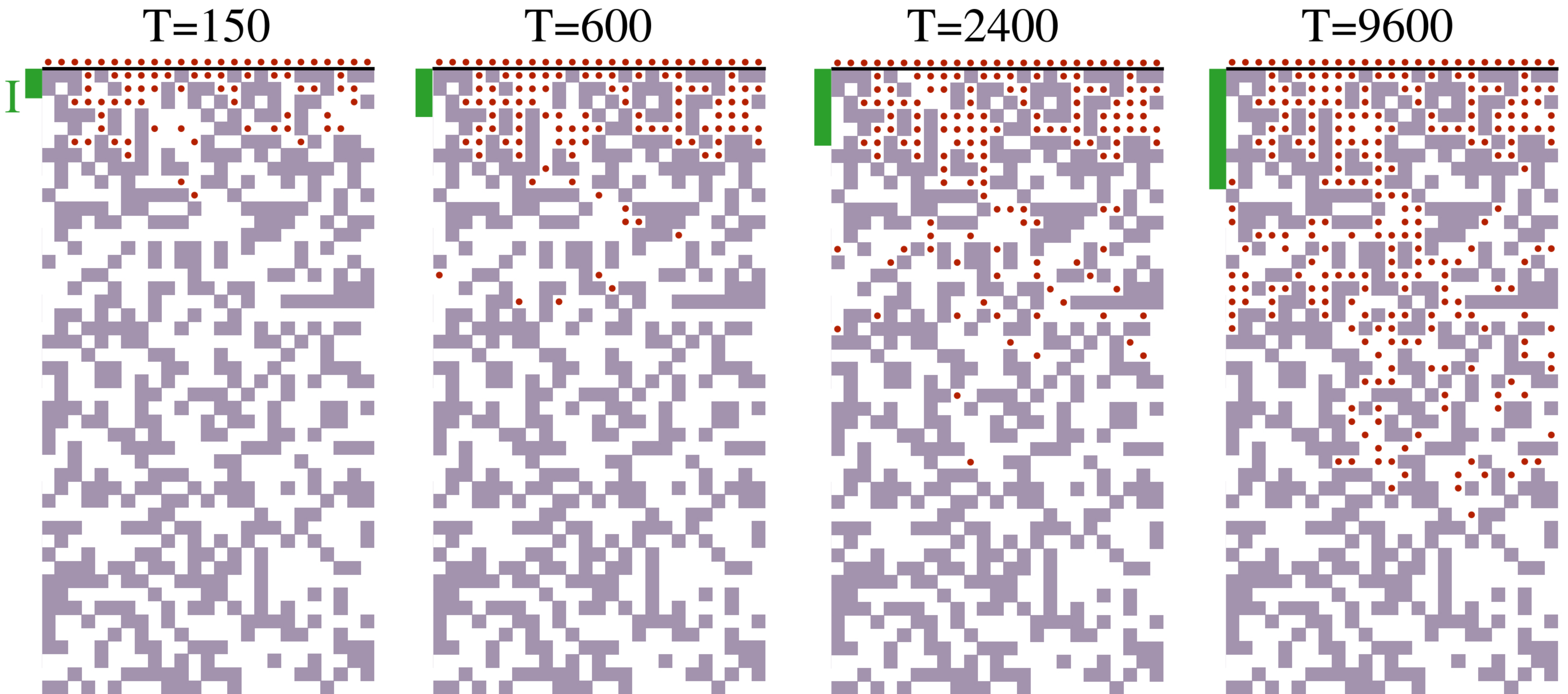}
\caption{(Color online)
Infiltration without reactions in a porous medium at the percolation threshold in $d=2$.
The vertical green bar represents the infiltration length.
}
\label{imageperc2d}
\end{figure*}

Figure \ref{imageperc2d} shows snapshots of an infiltrated medium in $d=2$ at several times.
The anomalous advance of the infiltration front is confirmed by the infiltration length bars:
as the time increases by a factor $4$ between consecutive panels, $I$ varies by
factors smaller than $2$ (the factor $2$ is expected in Fickean infiltration).
In $d=3$, $I$ varies by smaller factors when results at the same time are compared.
As the infiltration advances, we also observe a smaller density of filled pore branches,
which is related to the fractality of the percolation cluster.

In Ref. \protect\cite{reis2016}, a scaling approach predicted that 
the infiltration length in fractal porous media scales as in Eq. (\ref{defngeral}) with
\begin{equation}
n=\frac{d_F-d_B}{d_W} .
\label{Iperc}
\end{equation}
Eq. (\ref{Iperc}) was confirmed in several infinitely ramified fractals, including some cases with
superdiffusive infiltration ($1/2<n<1$) \citep{reisbolstervoller,reisvoller2019}.
However, it was not formerly tested in stochastic fractals.

\begin{figure}[!ht]
\center
\includegraphics[clip,width=0.37\textwidth]{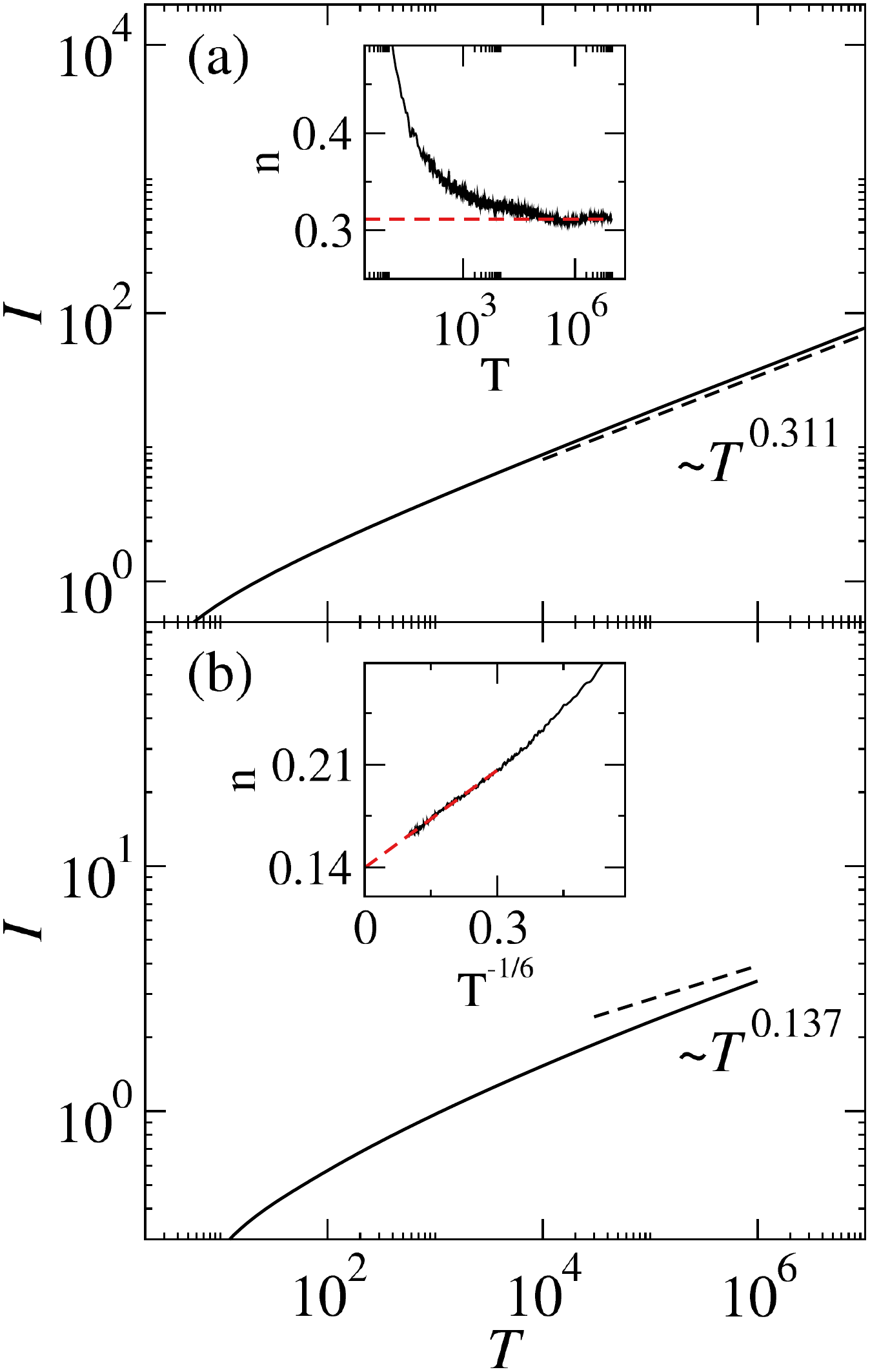}
\caption{(Color online)
$I$ as a function of $T$ in the infiltration without reactions in media at the
critical percolation point in (a) $d=2$ and (b) $d=3$.
The insets show the convergence of the effective exponents;
the variable in the abscissa of the inset in (b) was chosen to provide
the best linear fit of the data at the longest simulation times.
}
\label{infilt}
\end{figure}

Using the structural and dynamical exponents of percolation, Eq. (\ref{Iperc}) predicts
$n\approx 0.3113$ in $d=2$ and $n\approx 0.137$ in $d=3$.
Figs. \ref{infilt}(a) and \ref{infilt}(b) show the evolutions of the infiltration length
in percolating media in $d=2$ and $d=3$, respectively, with dashed lines indicating the
predicted exponents $n$.
The agreement in $d=2$ is very good, but deviations are observed in $d=3$ until the longest
simulated times.
The local slopes of the $\log{I}\times\log{T}$ plots are the effective infiltration exponents
$n_{eff}$ shown in the insets of Figs. \ref{infilt}(a) and \ref{infilt}(b).
In $d=2$, $n_{eff}$ reaches values very close to the prediction of Eq. (\ref{Iperc})
at $T\sim{10}^5$.
In $d=3$, $n_{eff}$ is more than $10\%$ larger than the predicted value at $T\sim{10}^6$.
The fit of $n_{eff}$ as a function of $T^{-1/6}$ leads to an asymptotic estimate consistent with
that of Eq. (\ref{Iperc}), which shows the presence of large corrections to the dominant scaling.

Note that the right panel in Fig. \ref{imageperc2d} shows that the distances of several S particles
from the source are much larger than $I$.
This occurs because their displacement is governed by Eq. (\ref{defdW}), with exponent $1/d_W$,
while $I$ increases with a smaller exponent $n$
[Eq. (\ref{Iperc}) with $d_F-d_B<1$ implies $n<1/d_W$].

\section{Simulations of Infiltration with Reactions}
\label{simulations}

\subsection{Results in $d=2$}
\label{2d}

\begin{figure*}[!ht]
\center
\includegraphics[clip,width=0.9\textwidth]{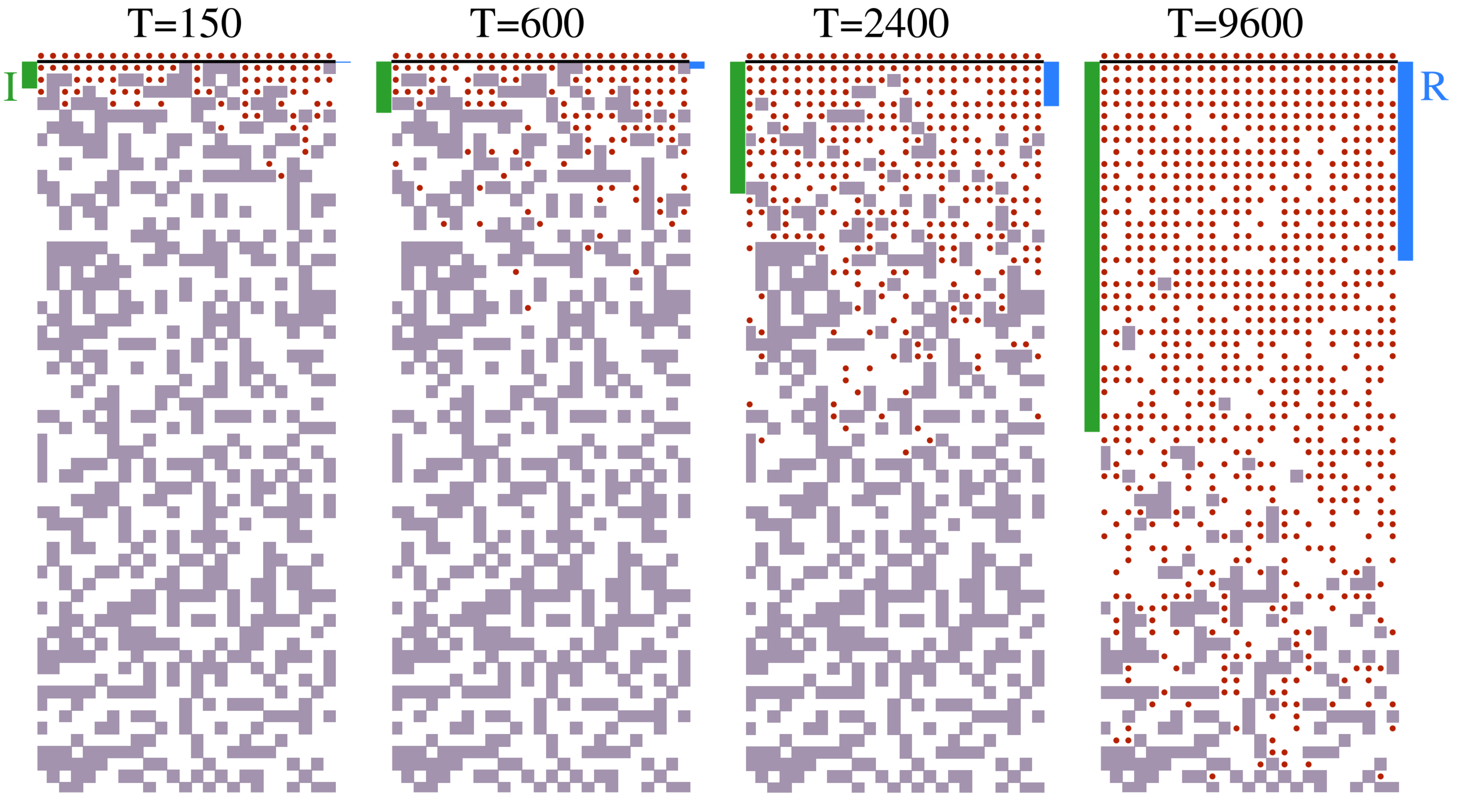}
\caption{(Color online)
Infiltration with dissolution ($r={10}^{-3}$)
in a porous medium at the percolation threshold in $d=2$.
In each panel, the left green bar represents the infiltration length and
the right blue bar represents the reaction length.
}
\label{imagedis2d}
\end{figure*}

Figure \ref{imagedis2d} illustrates the infiltration in $d=2$ for $r={10}^{-3}$ and
the same times of Fig. \ref{imageperc2d} (the case without reactions).
The results for $T<{10}^3$ show a slow advance of the infiltration front
because the reaction probability $\sim rT$ is small; this is similar to the case without reactions.
However, at longer times, the S particles fill some dissolved
(or partly dissolved) layers near the source.
The advance of the infiltration is faster than that without reactions.
The reaction length $R$ increases even faster.
If the results at $T=9600$ are compared with the case of a compact medium (Fig. \ref{compact_perfil}),
here we observe a deeper infiltration despite the smaller value of $r$.
This is possible because the solute infiltrates into the channels of the porous medium
and dissolves the channel walls, which facilitates the subsequent infiltration.

\begin{figure}[!ht]
\center
\includegraphics[clip,width=0.4\textwidth]{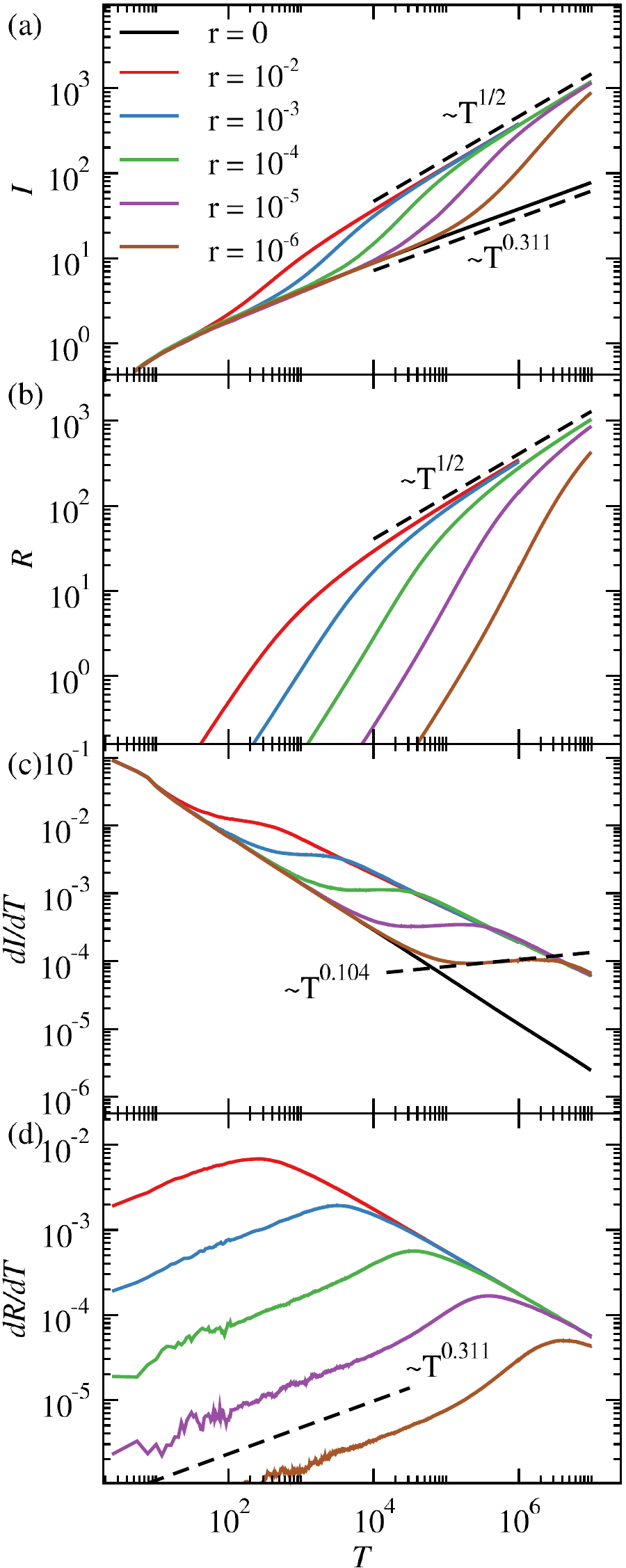}
\caption{
Evolution of (a) infiltration length and (b) reaction length in $d=2$ for several values of $r$;
the dashed lines are theoretical predictions for the subdiffusive and the Fickean regime.
(c) and (d) shows the respective rates of change, with dashed lines indicating the slopes theoretically
predicted for the subdiffusive and the intermediate regimes.
}
\label{infdis2d}
\end{figure}

Figures \ref{infdis2d}(a) and \ref{infdis2d}(b) show the time evolution of the infiltration length
and of the reaction length, respectively, for several values of $r$.
At short times, the scaling of $I$ is subdiffusive, similarly to the case without reactions;
in this regime, $R$ is very small, but rapidly increases in time.
The extent of the reactions eventually increase and lead to deviations from the subdiffusive scaling.
The time for these deviations to appear increases as $r$ decreases;
inspection of Figs. \ref{infdis2d}(a) and \ref{infdis2d}(b) shows that they occur
as $R\sim1$ in all cases.
Subsequently, rapid increases of $I$ and $R$ are observed, followed by the convergence of both 
quantities to the Fickean behavior $\sim T^{1/2}$ (the same asymptotic behavior observed
with an initially compact medium; Sec. \ref{review}).

Figures \ref{infdis2d}(c) and \ref{infdis2d}(d) show the evolution of the infiltration rate
$\dot{I}\equiv{\text d}I/{\text d}T$ and reaction rate $\dot{R}\equiv{\text d}R/{\text d}T$, respectively.
$\dot{I}$ decreases at short times, has a local minimum just after the subdiffusive
regime, and has a local maximum just before the Fickean regime.
Between those extrema, $\dot{I}$ slowly increases in time;
the time interval of this intermediate regime increases as $r$ decreases.
$\dot{R}$ increases as a power law in the subdiffusive regime of $I$ and
subsequently shows a faster time increase (which can hardly be fit as a power law);
at longer times, it also shows a maximum before the crossover to the Fickean regime.

The crossover time between the subdiffusive infiltration and this intermediate regime is denoted
as $T_1$ and the crossover time to the Fickean regime is denoted as $T_2$.
Here we consider two possible definitions of $T_1$:
in the first one, it is the time of the local
minimum of $\dot{I}$ after the subdiffusive scaling;
in the second one, it is the time in which the relative deviation of $I$
from the non-reactive case ($r=0$) reaches a preset value $\Delta$ (this is the same definition used in
Ref. \protect\cite{reisAWR2019} for a similar model in regular fractals).
The crossover time $T_2$ is also estimated considering two definitions:
in the first one, it is the time of the local
maximum of $\dot{I}$ before the Fickean regime;
in the second one, it is the time in which the relative deviation of $I$
from the asymptotic Fickean scaling reaches a preset value $\Delta$.
The latter considers the exact asymptotic relation $I=2\sqrt{Dt/(d\pi)}$ for
a lattice with only P sites.

\begin{figure}[!ht]
\center
\includegraphics[clip,width=0.32\textwidth]{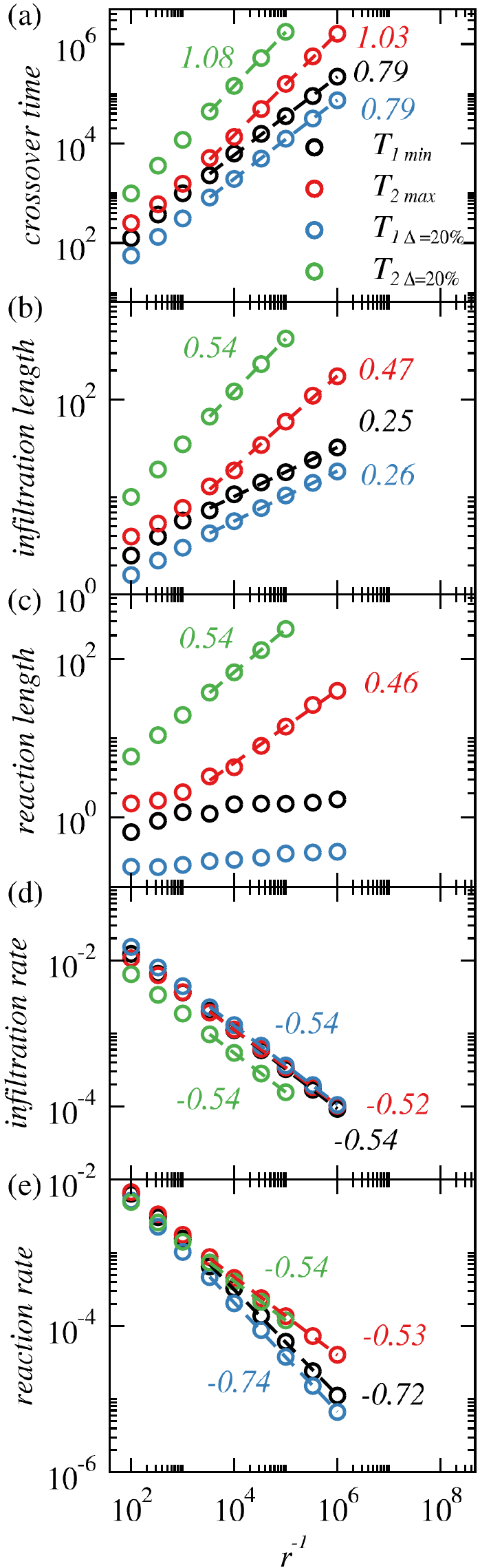}
\caption{(Color online)
(a) Crossover times $T_1$ and $T_2$ as a function of $r^{-1}$, in $d=2$, obtained from local extrema
of $I$ (labels $min$ and $max$) and from the deviation $\Delta=20\%$.
(b) Infiltration lengths ($I_1$, $I_2$) and (c) reaction lengths ($R_1$, $R_2$)
at the crossover times [same color code as (a)].
(d) Infiltration rates (${\dot{I}}_1$, ${\dot{I}}_2$) and (e) reaction rates
(${\dot{R}}_1$, ${\dot{R}}_2$) at the crossover times [same color code as (a)].
In all plots, the error bars are smaller than the sizes of the symbols and
the slopes of the least squares fits of all data sets (dashed lines) are indicated.
}
\label{crossovertimes2d}
\end{figure}

Fig. \ref{crossovertimes2d}(a) shows the crossover times calculated with the two definitions
as a function of $r^{-1}$.
In this analysis, we used $\Delta=20\%$.
The plots suggest power law scalings as
\begin{equation}
T_1\sim r^{-b}\qquad,\qquad T_2\sim r^{-c} ,
\label{t1t2}
\end{equation}
where $b$ and $c$ are positive exponents.
Linear fits of the $T_1$ data for $r\leq{10}^{-3}$ give $b=0.79$ with the two methods;
the fits of the $T_2$ data give $c=1.03$ and $1.08$.
Since $c$ is unequivocally larger than $b$, the ratio $T_2/T_1$ increases as $r$ decreases, i.e.
the time interval of the intermediate regime increases.
It may be very long for slow reactions;
for instance, for $r={10}^{-6}$, $T_2$ and $T_1$ differ by approximately $1.5$ order of magnitude.

The values of $I$ at $T_1$ and $T_2$, which we denote as $I_1$ and $I_2$, respectively,
are shown in Fig. \ref{crossovertimes2d}(b) as a function of $r^{-1}$.
As $r$ decreases, $I_1$ increases, which is consistent with a longer subdiffusive regime;
$I_2$ increases slightly faster, which means that a relatively larger infiltration
is obtained in the regime with time increasing rates.
The values of $R$ at those crossovers, which we denote as $R_1$ and $R_2$, respectively,
are shown in Fig. \ref{crossovertimes2d}(c).
$R_1\sim 1$ for all values of $r$, which suggests that this condition is related to the breakdown
of the subdiffusion.
In Fig. \ref{crossovertimes2d}(d), we show the infiltration rates $\dot{I}_1$ and $\dot{I}_2$
at the crossovers, and in Fig. \ref{crossovertimes2d}(e) we show the corresponding reaction rates
$\dot{R}_1$ and $\dot{R}_2$.
In all cases, a decrease of these rates is observed as $r$ decreases
(note that they are rates calculated at different times, so it does not contradict the observation
of time increasing rates in the intermediate regime for a constant $r$).
In Figs. \ref{crossovertimes2d}(b)-(e), the linear fits show that those
quantities vary as power laws of $r$ for small values of this parameter;
the exponents shown in the plots  
weakly depend on the definition used to calculate the crossover times.

\subsection{Results in $d=3$}
\label{3d}

\begin{figure*}[!ht]
\center
\includegraphics[clip,width=0.9\textwidth]{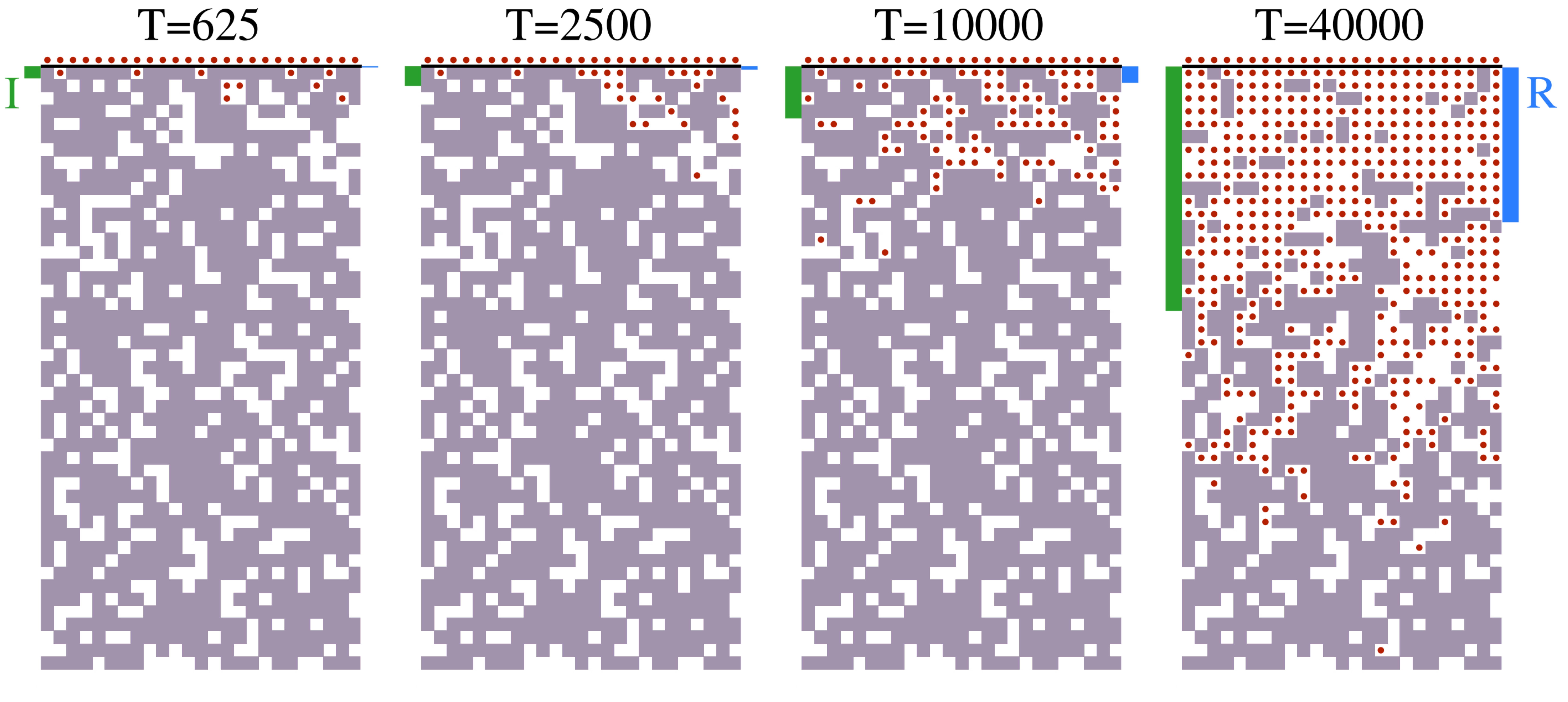}
\caption{Evolution of a small cross-section of the lattice in $d=3$ with $r=10^{-4}$.
The time $625$ is bellow $T_1$, $2500$ is close to $T_1$,
$10000$ is close to the inflection point, and $40000$ is close to $T_2$.
The left green bar represents the infiltration length and the right blue bar the reaction length.
}
\label{snapshots}
\end{figure*}

Fig. \ref{snapshots} shows cross sections of a medium during its infiltration
with reaction probability $r={10}^{-4}$.
In the first two panels, the infiltration is slow and a small number of M sites has reacted.
The third panel shows a high filling of the accessible region near the source, but with no
significant change due to the reaction.
Between the third and the fourth panel, the infiltration and the reaction rapidly advance.
This is accompanied by significant dissolution of some layers near the source
(within the reaction length), where the density of S particles is high.
The layers within or slightly below the infiltration length are also highly filled,
but only a fraction of the M sites have reacted.

\begin{figure}[!ht]
\center
\includegraphics[clip,width=0.4\textwidth]{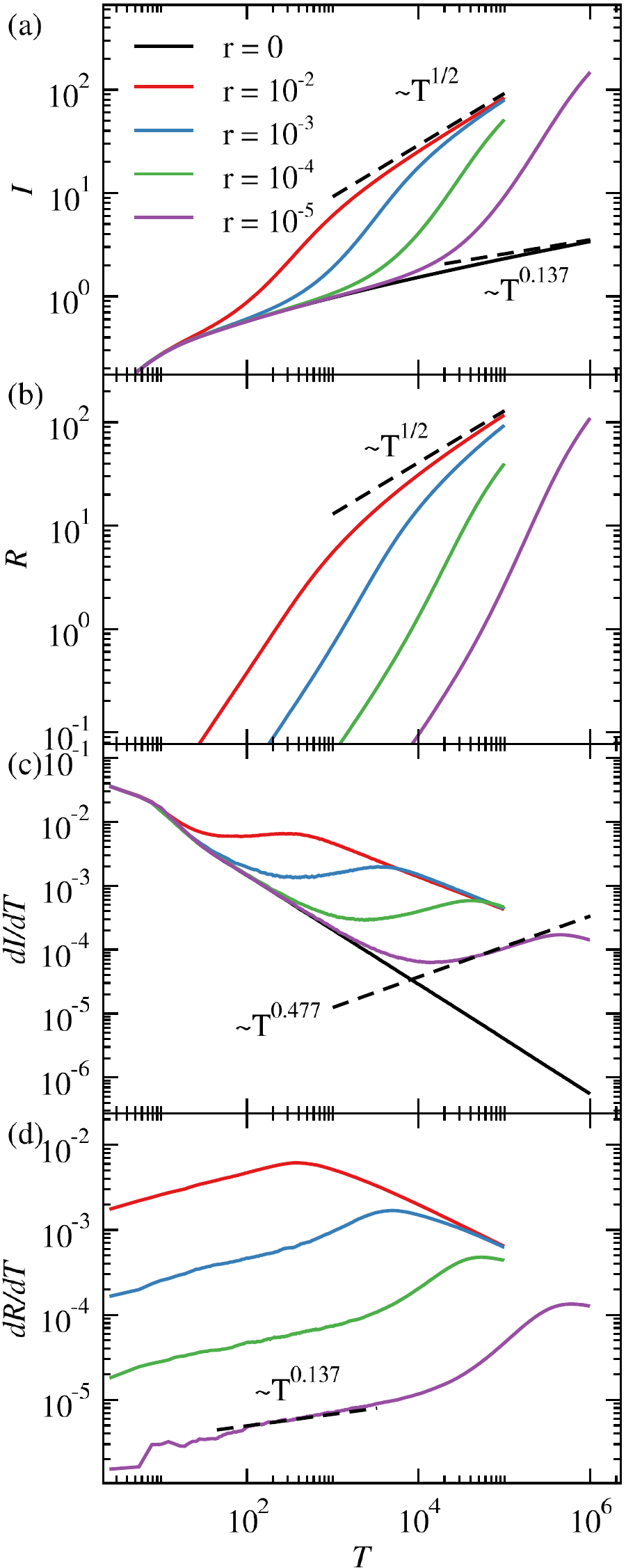}
\caption{
Evolution of (a) infiltration length and (b) reaction length in $d=3$ for several values of $r$;
the dashed lines are theoretical predictions for the subdiffusive and Fickean regime.
(c) and (d) shows the respective rates of change, with dashed lines indicating the slopes theoretically
predicted for the subdiffusive and intermediate regimes.
}
\label{infdis3d}
\end{figure}

Figures \ref{infdis3d}(a) and \ref{infdis3d}(b) show the evolution of $I$ and $R$,
respectively, for several $r$, and
Figs. \ref{infdis3d}(c) and \ref{infdis3d}(d) show the corresponding rates.
The qualitative evolution is similar to that in $d=2$, with an initial subdiffusive regime,
an intermediate regime with time increasing rates, and a final Fickean regime.
However, the distinct features of the intermediate regime are shaper in $d=3$:
the time increase of $\dot{I}$ is clearer and the increase of $\dot{R}$ is faster
in comparison with $d=2$.
Moreover, this regime lasts longer as $r$ decreases.

\begin{figure}[!ht]
\center
\includegraphics[clip,width=0.32\textwidth]{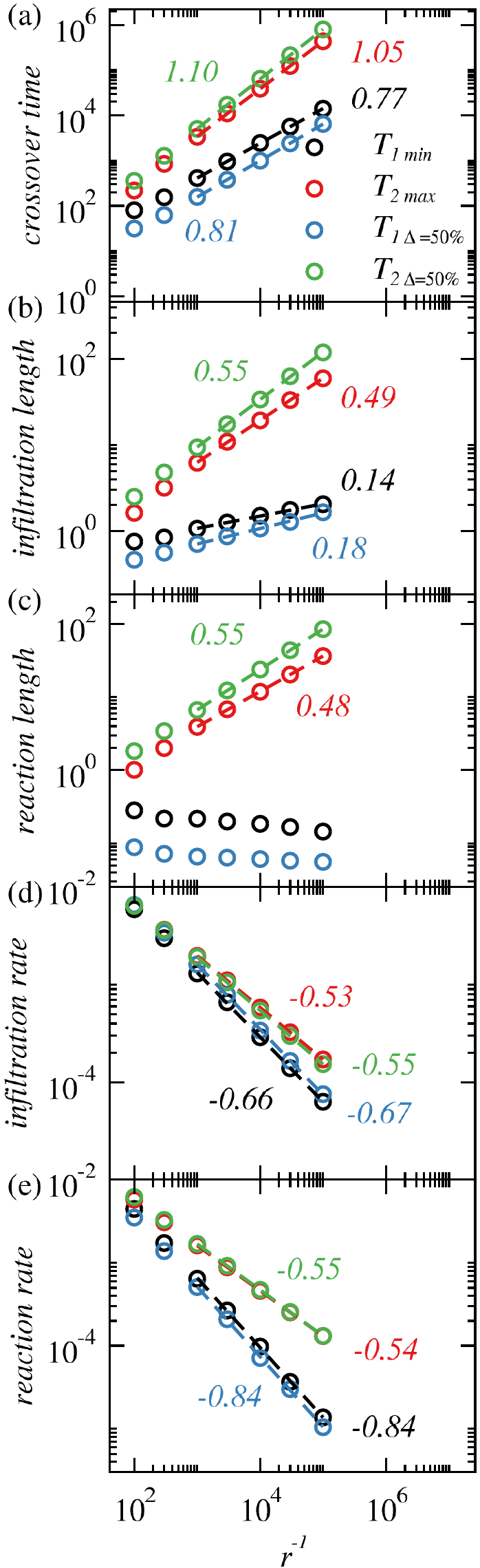}
\caption{
(a) Crossover times $T_1$ and $T_2$ as a function of $r^{-1}$, in $d=3$, obtained from local extrema
of $I$ (labels $min$ and $max$) and from the deviation $\Delta=50\%$.
(b) Infiltration and (c) reaction lengths at the crossover times [same color as (a)].
(d) Infiltration and (e) reaction rates at the crossover times [same color as (a)].
In all plots, the error bars are smaller than the sizes of the symbols and
the slopes of the least squares fits of all data sets (dashed lines) are indicated.
}
\label{crossovertimes}
\end{figure}

Here we also characterize the crossovers between the three scaling regimes by the times
$T_1$ and $T_2$, using the two definitions given in Sec. \ref{2d}.
For the method that considers fractional deviations of $I$ from the subdiffusive and from the Fickean
scalings, here we use $\Delta=50\%$.
The crossover times are shown in Fig. \ref{crossovertimes}(a) as a function of $r^{-1}$ and also
follow the power law relations of Eq. (\ref{t1t2}).
The exponents for $T_1$ are $b=0.77$ and $0.81$, while the exponents for $T_2$ are
$c=1.05$ and $1.10$; thus, the different definitions of those times also
lead to small differences in the measured exponents.
Since $c>b$, the number of time decades of the intermediate regime
[$\log_{10}{\left(T_2/T_1\right)}$] increases as $r$ decreases.

The infiltration lengths at $T_1$ and $T_2$ are shown in 
Figure \ref{crossovertimes}(b) as a function of $r^{-1}$;
the reaction lenghts at the crossover times are shown in Fig. \ref{crossovertimes}(c).
As in the two-dimensional case, here we
observe that $I_1$ increases as $r$ decreases, consistently with a longer subdiffusive
regime, whereas $I_2$ increases faster, consistently with larger infiltrations during the
intermediate regime of time increasing rates.
We also observe that $R_1\sim 1$ for all $r$, supporting the assumption that this condition
is related to the breakdown of the subdiffusion.
Figure \ref{crossovertimes}(d) shows the infiltration rates and
Fig. \ref{crossovertimes}(e) shows the reaction rates at the crossover times $T_1$ and $T_2$,
which follow similar trends as the crossover rates in $d=2$.
In Figs. \ref{crossovertimes}(b)-(e), the scaling exponents obtained from data fits with
the smallest $r$ are shown in the plots and are weakly dependent on the definition of
the crossover times.

\newpage

\section{Scaling Approach}
\label{scaling}

For any reaction probability $r$, the short time subdiffusive infiltration is described by
Eq. (\ref{Iperc}) because the number of dissolved M sites is small.
Reactions may occur on the walls of the infiltrated channels, i.e. on the faces of the M sites
in contact with S particles.
An analogy with the dissolution of the compact medium (Sec. \ref{review}) is helpful here:
at short times, Figs. \ref{compact_perfil}(a)-(c) show that almost all P sites in contact
with the M sites are infiltrated.
Thus, the dissolved region acts like an extension of the source at times much shorter
than the crossover to the Fickean behavior.
Here, the difference is that the S particles also invade the original P sites of the porous medium,
i.e. the extended source is inside the porous medium.

This reasoning implies that the number of faces of M sites in contact with S particles
is proportional to the number $N_I$ of infiltrated P sites.
In $d=3$, the area of M sites in contact with S particles is of order
$A_I\sim N_Ia^2\sim IL^2$;
in $d=2$, the length in contact with S particles is $L_I\sim N_I a\sim IL$.
These relations omit the average number of faces of M sites
per infiltrated P site, which is of order $1$.

In the neighborhood of an infiltrated P site, an M site reacts with probability $r$ in a time $\tau$.
Thus, in a time $t$, the reaction advances into an average length
$l_\bot\sim a\left(r/\tau\right)t=arT$ of the M sites surrounding that P site.
This is applicable in $d=2$ and $d=3$ at short times, in which $l_\bot\lesssim a$,
i.e. in which the advance of the reaction is limited to the M sites in the closest neighborhood
of the P site.
The total volume of M that reacts in a time $T$ is $V_R\sim A_Il_\bot\sim IaL^2rT$ in $d=3$;
in $d=2$, the total area that reacts is $A_R\sim L_Il_\bot\sim IaLrT$.
Using Eqs. (\ref{defR}) and (\ref{Iperc}), the reaction length in both dimensions scales as
\begin{equation}
R\sim rT^{n+1} .
\label{Rperc}
\end{equation}
Since $n>0$, $R$ increases faster than linearly since short times.
The corresponding rate of reaction is
\begin{equation}
\frac{{\text d}R}{{\text d}t}\sim rT^n .
\label{dRdt}
\end{equation}
In $d=2$, the plots in Fig. \ref{infdis2d}(d) have slopes consistent with the estimate
$n\approx 0.3113$.
In $d=3$, the plots in Fig. \ref{infdis3d}(d) have slopes slightly larger
than the estimate $n\approx 0.137$; however, this is expected because the effective exponents
$n_{eff}$ are large at short times, as shown in the inset of Fig. \ref{infilt}(b).

After some time, several M sites are transformed into P sites, which
facilitates the infiltration of other S particles.
Figs. \ref{crossovertimes2d}(c) and \ref{crossovertimes}(c) show that the first crossover occurs
when $R\sim 1$, which corresponds to the reaction of one M site per source site.
This means that the extended source formed inside the porous medium has a volume
of the same order as the volume of the source localized at the infiltration border.
After it occurs, the attack to the M sites inside the medium (by an increasing number of S particles)
is faster than the attack to the top M sites;
the subdiffusive behavior then ceases and a different scaling regime begins.
The crossover time $T_1$ is obtained by substituting the condition $R\sim1$ in Eq. (\ref{Rperc}),
which leads to the scaling in Eq. (\ref{t1t2}) with exponent
\begin{equation}
b=\frac{1}{1+n} .
\label{scalingt1}
\end{equation}
The infiltration and reaction lengths at $T_1$ are obtained by substituting Eqs. (\ref{t1t2})
and (\ref{scalingt1}) in Eqs. (\ref{Iperc}) and (\ref{Rperc}); their time derivatives
give the rates of those lengths at the crossover:
\begin{equation}
I_1\sim r^{-n/\left( n+1\right)} \qquad , \qquad {\dot{I}}_1\sim r^{\left( 1-n\right)/\left( 1+n\right)} 
\qquad , \qquad {\dot{R}}_1\sim r^{b} .
\label{scalingI1}
\end{equation}

Our numerical results support these scaling relations.
In $d=2$, the estimate $n\approx 0.3113$ gives $b\approx0.763$, $n/\left( n+1\right)\approx0.237$,
and $\left( 1-n\right)/\left( 1+n\right)\approx0.525$ for the exponents in Eq. (\ref{scalingI1}).
The numerical values of the exponents of $T_1$, $I_1$, ${\dot{I}}_1$, and ${\dot{R}}_1$,
shown in Figs. \ref{crossovertimes2d}(a)-(c), are close to the predictions of this
scaling approach, with maximal deviations $\approx 6\%$.
In $d=3$, the asymptotic value $n\approx 0.137$ gives
$b\approx0.880$, $n/\left( n+1\right)\approx0.120$, and $\left( 1-n\right)/\left( 1+n\right)\approx0.759$.
The numerical estimates of the exponents of $T_1$,
${\dot{I}}_1$, and ${\dot{R}}_1$ [Figs. \ref{crossovertimes}(a)-(c)] differ up to
$\approx 13\%$ from those values.
The numerical estimate of the exponent of $I_1$ [Fig. \ref{crossovertimes}(b)] has a larger deviation.
In this case, $T_1$ is in the range ${10}^{2}{\text{--}}{10}^{4}$ [Fig. \ref{crossovertimes}(a)],
in which the infiltration length without reactions increase with $n_{eff}$ between $0.25$ and $0.18$
[Fig. \ref{infilt}(b)], which are $30\%{\text{--}}80\%$ larger than the asymptotic value $\approx 0.137$.
Thus, the numerical estimates of exponents $b$ and $\left( 1-n\right)/\left( 1+n\right)$ are actually
expected to be smaller than the scaling predictions based on the asymptotic $n$,
whereas the estimate of $n/\left( n+1\right)$ is expected to be larger than the scaling prediction.

\begin{figure}[!ht]
\center
\includegraphics[clip,width=0.4\textwidth]{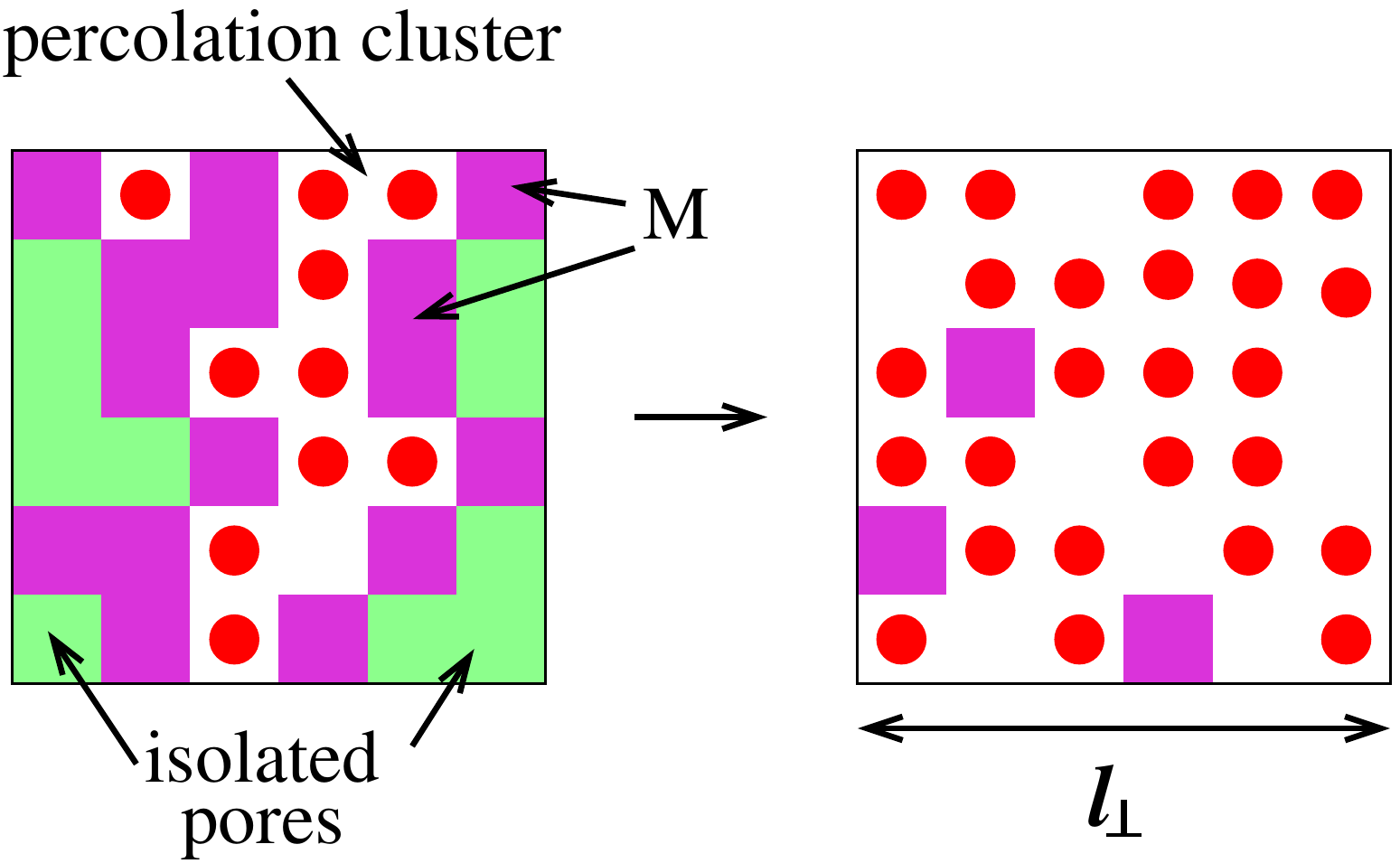}
\caption{(Color online)
Scheme of an infiltrated region of the porous medium in $d=2$ before (left) and after (right)
the reactions occur.
}
\label{region}
\end{figure}

The deviation from subdiffusion at $T_1$ implies that the fractality is broken in the region within
the infiltration length.
At $T>T_1$, the dissolution of M sites creates new paths for the infiltration.
The S particles fill P sites that belong to the critical percolation cluster plus
new P sites created by the reactions and P sites that were initially isolated;
see Fig. \ref{region}.
To estimate the infiltration length, we consider these contributions separately.

First consider the infiltrated part of the original percolation cluster.
The infiltration source is extended, i.e. the non-infiltrated parts of that cluster are
below a region with a high density of S particles.
Thus, the infiltration advance is expected to be the same observed
when the initial percolation cluster was in contact with the source at $z=0$.
In each time interval $T_1$, the infiltration advances into that cluster by a length $I_1$
in the $z$ direction (as in the initial subdiffusive regime), which means that the
extended source advances by that length.
At $T>T_1$, the infiltration length $I_{pc}$ in the original percolation cluster is proportional
to $T/T_1$:
\begin{equation}
I_{pc}\sim I_1 \frac{T}{T_1} .
\label{Io}
\end{equation}

Now we analyze the infiltration of the other P sites, i.e. sites that do not belong
to the original percolation cluster.
The reactions around an infiltrated P site advance in all directions
to a length $l_\bot\sim a\left(r/\tau\right)t=arT$, as illustrated in Fig. \ref{region};
now $l_\bot$ is not of the same order as $a$.
In a $d$-dimensional region with size $l_\bot$, the total number of P sites is
$\sim p_c{\left(l_\bot/a\right)}^d$, i.e. the P sites occupy a finite fraction of the available volume.
In the same region, the number of P sites in the critical percolation cluster is
$\sim{\left(l_\bot/a\right)}^{d_F}$.
Thus, the ratio between the total number of P sites and
the number of P sites in the percolation cluster at time $T$ is
\begin{equation}
f_P\left( T\right) \sim{\left(\frac{l_\bot}{a}\right)}^{d-d_F} \sim {\left( rT\right)}^{d-d_F} .
\label{fP}
\end{equation}
Since $d>d_F$, this result implies that most of the infiltrated sites are not those of the 
original percolation cluster.

Recalling that $I_{pc}$ [Eq. (\ref{Io})] accounts only for the infiltrated sites
of the percolation cluster, the total infiltrated length $I$ is larger than $I_{pc}$
by the factor $f_P\left( T\right)/f_P\left( T_1\right)$:
\begin{equation}
I\sim \frac{f_P\left( T\right)}{f_P\left( T_1\right)} I_{pc}\sim r^h t^g \qquad , \qquad
g\equiv 1+d-d_F \qquad , \qquad h\equiv\frac{g-n}{n+1} .
\label{scalingIsuper}
\end{equation}
This gives $I\sim I_1$ at $T\sim T_1$.
The infiltration rate scales as
\begin{equation}
\dot{I} \sim r^h T^{d-d_F} .
\label{scalingdIdTsuper}
\end{equation}

Eqs. (\ref{scalingIsuper}) and (\ref{scalingdIdTsuper}) give $g>1$ in any dimension, so they
theoretically predict the time increasing infiltration rate.
The numerical values are $g=53/48\approx 1.104$ in $d=2$ and $g\approx1.477$ in $d=3$,
which show that the effect is more pronounced in $d=3$.
This result is confirmed with good accuracy by the slopes $d-d_F\approx0.104$ and $0.477$
of the infiltration rate evolution in Figs. \ref{infdis2d}(c) and \ref{infdis3d}(c), respectively.

The crossover to Fickean infiltration occurs when the infiltration length of
Eq. (\ref{scalingIsuper}) is of the same order as the diffusive infiltration length $I\sim T^{1/2}$,
which does not depend on $r$.
This gives the scaling of $T_2$ as in Eq. (\ref{t1t2}) with
\begin{equation}
c=\frac{g-n}{\left(1+n\right)\left(g-1/2\right)} .
\label{scalingt2}
\end{equation}
At the crossover, the infiltration length and its rate scale as
\begin{equation}
I_2\sim r^{-c/2} \qquad , \qquad {\dot{I}}_1\sim r^{c/2} .
\label{scalingI2}
\end{equation}

In $d=2$, Eq. (\ref{scalingt2}) gives $c\approx 1.01$ ($c/2\approx0.51$).
The numerical estimates of $c$ obtained from $T_2$ [Fig. \ref{crossovertimes2d}(a)]
and the estimates of $c/2$ obtained from $I_2$ and ${\dot{I}}_2$ 
[Figs. \ref{crossovertimes2d}(d) and \ref{crossovertimes2d}(e)] are close to these predictions.
In $d=3$, Eq. (\ref{scalingt2}) gives $c\approx 1.21$ ($c/2\approx0.61$).
However, the same relation with $n_{eff}$ larger than the asymptotic $n$,
as shown in Fig. \ref{infilt}(b), gives a smaller effetive value of $c$.
Indeed, the numerical estimates of exponent $c$ in the scaling of $T_2$ [Fig. \ref{crossovertimes}(a)]
are smaller than the estimate of the scaling approach and the numerical estimates of the exponents
of $I_2$ and of ${\dot{I}}_2$ [Figs. \ref{crossovertimes}(b) and \ref{crossovertimes}(c)]
are smaller than the corresponding estimate of $c/2$,
with differences in the range $9{\text{--}}13\%$.

The number of time decades $N_{tir}$ with time increasing rates is 
\begin{equation}
N_{tir}\sim\log_{10}{\left(\frac{T_2}{T_1}\right)}\sim\left(c-b\right)\log_{10}{r^{-1}} ,
\label{scalingTsb}
\end{equation}
where $c-b\approx 0.25$ in $d=2$ and $c-b\approx 0.33$ in $d=3$.
In cases of very slow reactions or very fast diffusion, Eq. (\ref{pkD}) implies $r\ll1$
and the time increasing rates may be observed in several time decades.

For comparison, in the reactive regime of the dissolution of a compact medium, the constant rate
lasts longer, with a number of time decades $\sim\log_{10}{r^{-2}}$.
However, in that case, the crossover to Fickean infiltration occurs with continuously decreasing
rates instead of the time increasing rates obtained here.
The difference is related to the penetration of the reactants in the percolating porous system,
which dissolve the pore walls and accelerate the convergence to the Fickean regime.

\begin{figure}[!ht]
\center
\includegraphics[clip,width=0.4\textwidth]{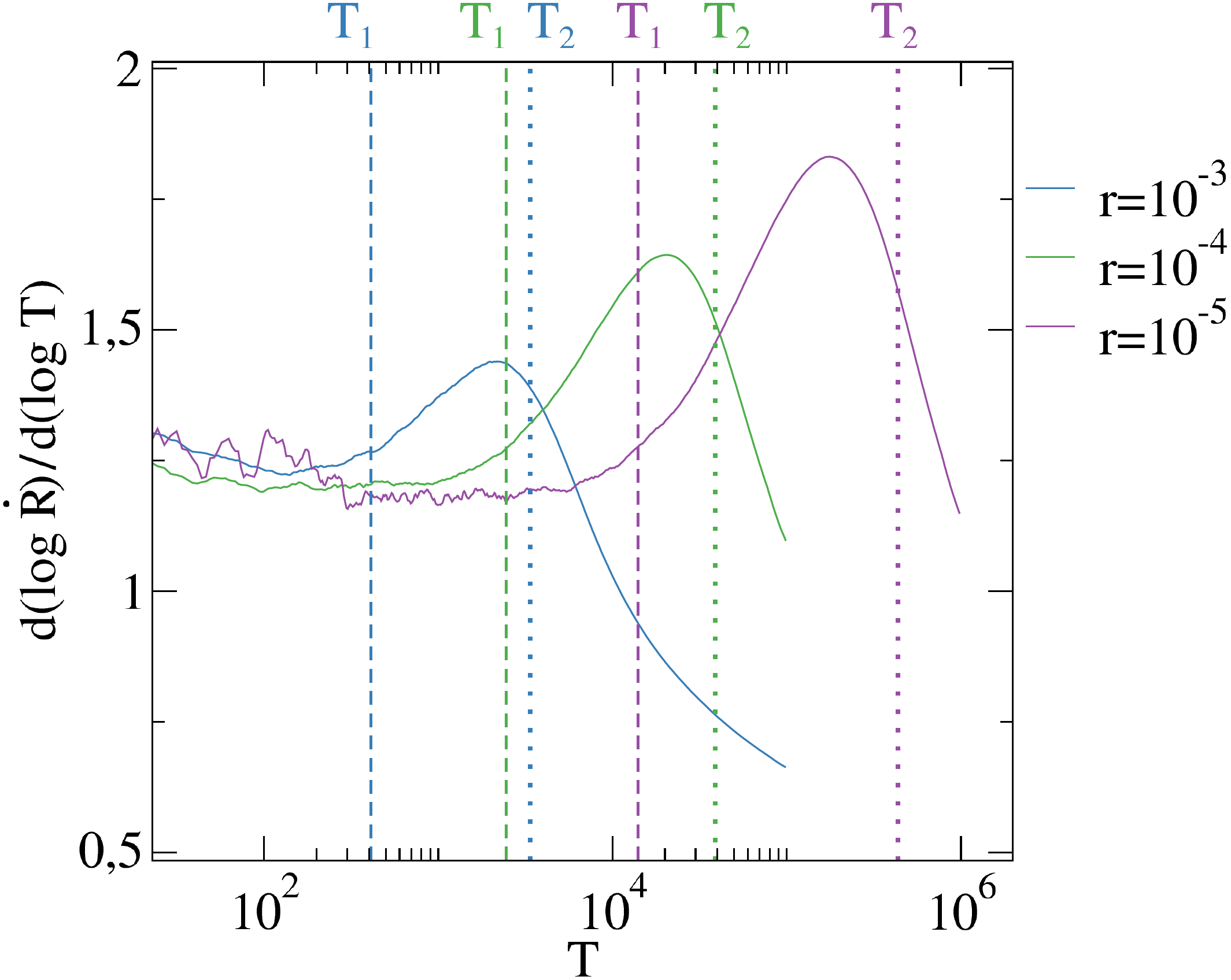}
\caption{ (Color online)
Effective exponent of $R$ as a function of $T$. The vertical dashed and dotted lines represent the crossover 
times $T_1$ and $T_2$ respectively, with colors matching the corresponding data set.}
\label{effectiveR}
\end{figure}

Our numerical results also show time increasing rates of the reaction lengths.
However, we were not able to determine a scaling relation for that length.
In Fig. \ref{effectiveR}, we show the effective slopes of the $\log{R}\times\log{T}$ plots for
three values of $r$ in $d=3$.
These slopes have peaks between $T_1$ and $T_2$, but they do not converge to a finite exponent
as $r$ decreases; instead, they are increasing, which means that faster growths of $R$ and
$\dot{R}$ are observed in the regime of time increasing infiltration rate.
Similar results are obtained in $d=2$.
Direct inspection of Figs. \ref{infdis2d}(b), \ref{infdis2d}(d), \ref{infdis3d}(b), and \ref{infdis3d}(d)
also show that it is not possible to perform reliable linear fits of the $\log{R}\times\log{T}$
or of the $\log{\dot{R}}\times\log{T}$ plots in that regime.
Despite this limitation, we observe that $R_2\sim I_2$ and
${\dot{R}}_2\sim{\dot{I}}_2$ in both dimensions,
which indicate that the reaction length crosses over to the Fickean regime simultaneously with
the infiltration length.

\section{Conclusion}
\label{conclusion}

We studied a model of solute infiltration from a source of constant concentration
into disordered lattices at the percolation thresholds and reactions of the solute
with the impermeable sites, which create new porous sites.
We performed numerical simulations in hypercubic lattices in $d=2$ and $3$ and
developed scaling approaches to determine the time evolution of the extents of infiltration
and reaction and to relate them with a model parameter $r$ that describes the relative
rate of reaction and diffusion (Damkohler number).
Cases of slow reactions ($r\ll1$) are considered.

The model without reactions shows subdiffusive infiltration with a time scaling exponent $n$
predicted by the same relation previously verified in infinitely ramified fractals.
In the reactive case, short time subdiffusion is observed and, at sufficiently long times,
the reacting media is far from the source and the infiltration is Fickean.
Between these regimes, a regime with time increasing rates of infiltration and reaction is observed.
This is explained by a cooperative effect between the directional infiltration of the solute
into the fractal porous medium and the advance of reactions in all directions.
The exponents of the time evolution of the infiltration and reaction lengths and their relations
with $r$ are predicted by a scaling approach and confirmed by numerical simulations, with
some deviations in $d=3$ that can be explained by the slow convergence of the initial subdiffusion
exponent.
The regime with time increasing rates spans a time range that increases as $r$ decreases,
so that it is more likely to be observed in slowly reacting materials.

The interplay of infiltration in porous media and chemical reactions
that change their structures is essential to understand their evolution.
This was already shown in the study of materials of geological and technological interest.
In low porosity systems, diffusion is expected to be the main transport mechanism, and
if those systems are fractal, they are expected to display subdiffusion.
This work suggests the investigation of a possible anomalously fast infiltration
if the reactions increase the porosity of such systems.

\begin{acknowledgments}

This work was supported by the Brazilian agencies CNPq (305391/2018-6),
FAPERJ (E-26/202.355/2018, E-26/202.881/2018, and E-26/210.354/2018),
and CAPES (88887.310427/2018-00).

\end{acknowledgments}

\bibliography{dissolution}

\end{document}